\documentclass[12pt,preprint]{aastex}

\newcommand{\lya}{Ly$\alpha$}
\newcommand{\lyb}{Ly$\beta$}
\newcommand{\lyg}{Ly$\gamma$}
\newcommand{\hi}{H~{\sc i}}
\newcommand{\oi}{O~{\sc i}}
\newcommand{\cii}{C~{\sc ii}}
\newcommand{\civ}{C~{\sc iv}}
\newcommand{\siii}{Si~{\sc ii}}
\newcommand{\siiv}{Si~{\sc iv}}
\newcommand{\mgi}{Mg~{\sc i}}
\newcommand{\mgii}{Mg~{\sc ii}}
\newcommand{\mhi}{{\rm H \; \mbox{\tiny I}}}
\newcommand{\moi}{{\rm O \; \mbox{\tiny I}}}
\newcommand{\msiii}{{\rm Si \; \mbox{\tiny II}}}
\newcommand{\msiiv}{{\rm Si \; \mbox{\tiny IV}}}
\newcommand{\mcii}{{\rm C \; \mbox{\tiny II}}}

\newcommand{\kms}{km~s$^{-1}$}
\newcommand{\kmsMpc}{km~s$^{-1}$~Mpc$^{-1}$}

\newcommand{\mOm}{\Omega_{\rm m}}
\newcommand{\mOl}{\Omega_{\Lambda}}

\newcommand{\dNdX}{\partial\mathcal{N}/\partial X}
\newcommand{\mdNdX}{\frac{\partial\mathcal{N}}{\partial X}}

\shorttitle{\oi\ at $z \sim 6$} 

\shortauthors{Becker et al.}

\begin{document}

\title{Discovery of excess \oi\ absorption towards the $z = 6.42$ QSO
  SDSS~J1148+5251\altaffilmark{1}}

\author{George D. Becker\altaffilmark{2}, Wallace L. W.
  Sargent\altaffilmark{2}, Michael Rauch\altaffilmark{3}, Robert
  A. Simcoe\altaffilmark{4}}

\altaffiltext{1}{The observations were made at the W.M. Keck
  Observatory which is operated as a scientific partnership between
  the California Institute of Technology and the University of
  California; it was made possible by the generous support of the
  W.M. Keck Foundation.}
\altaffiltext{2}{Palomar Observatory, California Institute of
  Technology, Pasadena, CA 91125, USA; gdb@astro.caltech.edu,
  wws@astro.caltech.edu}
\altaffiltext{3}{Carnegie Observatories, 813 Santa Barbara Street,
  Pasadena, CA 91101, USA; mr@ociw.edu}
\altaffiltext{4}{MIT Center for Space Research, 77 Massachusetts
  Ave. \#37-664B, Cambridge, MA 02139, USA; simcoe@mit.edu}

\begin{abstract}

  We present a search for \oi\ in the spectra of nine $4.9 \le z_{\rm
    QSO} \le 6.4$ QSOs taken with Keck/HIRES.  We detect six systems
  with $N_{\moi} > 10^{13.7}~{\rm cm^{-2}}$ in the redshift intervals
  where \oi~$\lambda 1302$ falls redward of the \lya\ forest.  Four of
  these lie towards SDSS~J1148+5251 ($z_{\rm QSO} = 6.42$).  This
  imbalance is unlikely to arise from variations in sensitivity among
  our data or from a statistical fluctuation.  The excess \oi\ occurs
  over a redshift interval that also contains transmission in \lya\
  and \lyb.  Therefore, if these \oi\ systems represent pockets of
  neutral gas, then they must occur within or near regions of the IGM
  that are highly ionized.  In contrast, no \oi\ is detected towards
  SDSS~J1030+0524 ($z_{\rm QSO} = 6.30$), whose spectrum shows
  complete absorption in \lya\ and \lyb\ over $\Delta z \sim 0.2$.
  Assuming no ionization corrections, we measure mean abundance ratios
  $\langle [{\rm O}/{\rm Si}] \rangle = -0.04 \pm 0.06$, $\langle
  [{\rm C}/{\rm O}] \rangle = -0.31 \pm 0.09$, and $\langle [{\rm
    C}/{\rm Si}] \rangle = -0.34 \pm 0.07$ ($2\sigma$), which are
  consistent with enrichment dominated by Type II supernovae.  The
  O/Si ratio limits the fraction of silicon in these systems
  contributed by metal-free very massive stars to $\lesssim 30\%$, a
  result which is insensitive to ionization corrections.  The ionic
  comoving mass densities along the $z_{\rm QSO} > 6.2$ sightlines,
  including only the detected systems, are $\Omega_{\moi} = (7.0 \pm
  0.6) \times 10^{-8}$, $\Omega_{\msiii} = (9.6 \pm 0.9) \times
  10^{-9}$, and $\Omega_{\mcii} = (1.5 \pm 0.2) \times 10^{-8}$.

\end{abstract}


\section{Introduction}

The state of the intergalactic medium (IGM) at redshift $z \sim6$
remains under considerable debate.  Significant transmitted flux in
the \lya\ forest at $z \sim 5.7$ means that the IGM must have been
highly ionized by at least $\sim 1$~Gyr after the Big Bang
\citep{becker01,djorgovski01,fan01,fan04}.  Each of the four known
QSOs at $z_{\rm QSO} \ge 6.2$ show Gunn-Peterson troughs \citep{gp65}
over at least a narrow redshift interval
\citep{becker01,pentericci02,fan01,fan03,fan04,white03}.  This
complete lack of transmitted flux has been interpreted as an
indication that the tail end of cosmic reionization may extend to $z
\sim 6$.  However, \citet{songaila04} found the evolution of
transmitted flux over $2 < z < 6.3$ to be consistent with a smoothly
decreasing ionization rate and not indicative of a sudden jump in the
\lya\ optical depth at $z \sim 6$.

Significant variations in the fraction of transmitted flux are common
among sightlines at the same redshift \citep{songaila04}.  While the
spectrum of SDSS~J1030+0524 ($z_{\rm QSO} = 6.30$) shows complete
absorption in \lya\ and \lyb\ over a redshift interval $\Delta z
\approx 0.2$ \citep{white03}, transmitted flux appears over the same
redshifts in the \lya, \lyb, and \lyg\ forests of SDSS~J1148+5251
($z_{\rm QSO} = 6.42$) \citep{white05,ohfur05}.  Either the IGM is
highly ionized everywhere at $z \sim 6.3$ and long stretches of
complete absorption are the result of line blending, or the neutral
fraction of the IGM is patchy on large scales.  A patchy IGM could
result from the clumpiness of the IGM and the clustering of ionizing
sources \citep{furoh05}.

Studies of \hi\ transmitted flux are ultimately hampered by the large
optical depths expected for even a small neutral fraction.  The
presence of a Gunn-Peterson trough can at best constrain the volume-
and mass-weighted \hi\ neutral fractions to $\gtrsim 10^{-3}$ and
$\gtrsim 10^{-2}$, respectively \citep{fan02}.  Alternative
measurements are required to probe larger neutral fractions.  The
non-evolution in the luminosity function of \lya-emitting galaxies
provides an independent indication that the IGM at $z \sim 6.5$ is
highly ionized \citep[e.g.,][]{malhotra04,stern05}, although \lya\
photons may escape as a result of galactic winds \citep{santos04} or
from locally ionized bubbles created by clustered sources
\citep{furlanetto04a}.  In the future, observations of redshifted
21~cm emission/absorption should trace the growth and evolution of
ionized regions at $z > 6$, placing strong constraints on reionization
scenarios \citep[e.g.,][]{tozzi00,kassim04,carilli04,furlanetto04b}.

A high neutral fraction must also have a measurable effect on metal
absorption lines \citep{oh02,furlanetto03}.  Overdense regions of the
IGM should be the first to become enriched, due to the presence of
star-forming sources, yet the last to remain ionized, due to the short
recombination times \citep{oh02}.  Low-ionization metal species should
therefore produce numerous absorption features in the spectra of
background objects prior to reionization.  A particularly good
candidate is \oi, which has an ionization potential nearly identical
to that of \hi\ and a transition at 1302~\AA\ that can be observed
redward of the \lya\ forest.  Oxygen and hydrogen will lock in
charge-exchange equilibrium \citep{osterbrock89}, which ensures that
their neutral fractions will remain nearly equal,
\begin{equation}
  \label{eq:equilib}
  f_{\mhi} \equiv \frac{n_{\mhi}}{n_{\rm H}} \approx \frac{n_{\moi}}{n_{\rm O}} \, .
\end{equation}
Despite the increased photo-ionization cross section of \oi\ at higher
energies, this relationship is expected to hold over a wide range in
$f_{\mhi}$ (Oh et al.~2005, in prep).

In this work, we present a search for \oi\ in the spectra of nine $4.9
\le z_{\rm QSO} \le 6.4$ QSOs.  This is the first time a set of
$z_{\rm QSO} > 5$ spectra has been taken at high resolution (R =
45,000).  Our sample includes three objects at $z_{\rm QSO} > 6.2$,
where we might expect to see an ``\oi\ forest'' \citep{oh02}.  In \S2
we describe the observations and data reduction.  The results of the
\oi\ search are detailed in \S3.  In \S4 we demonstrate a significant
overabundance of \oi\ systems towards the highest-redshift object,
SDSS~J1148+5251 ($z_{\rm QSO} = 6.42$), and compare the overabundance
to the number density of lower-redshift \oi\ systems and other
absorbers with high \hi\ column densities.  Measurements of the
relative metal abundances for all the detected systems are described
in \S5.  In \S6 we discuss the significance of these excess systems
for the enrichment and ionization state of the IGM at $z \sim 6$.  Our
results are summarized in \S7.  Throughout this paper we assume
$\Omega = 1$, $\mOm=0.3$, $\mOl = 0.7$, and $H_0 = 70$~\kmsMpc.

\section{The Data}

Observations using the Keck HIRES spectrograph \citep{vogt04} were
made between 2003 February and 2005 June, with the bulk of the data
acquired during 2005 January and February.  Our QSO sample and the
observations are summarized in Table~1.  All except the 2003 February
data were taken with the upgraded HIRES detector.  We used an
0\farcs86 slit, which gives a velocity resolution FWHM of $\Delta v =
6.7$~\kms.

The continuum luminosities in our sample approach the practical
detection limit of HIRES.  In addition, the spectral regions of
interest for this work lie in the far red, where skyline contamination
and atmospheric absorption are major concerns.  To address these
difficulties, the data were reduced using a custom set of IDL routines
written by one of us (GDB).  This package uses optimal sky subtraction
\citep{kelson03} to achieve Poisson-limited residuals in the
two-dimensional sky-subtracted frame.  One-dimensional spectra were
extracted using optimal extraction \citep{horne86}.  Blueward of the
\lya\ emission line, the continuum for these objects is highly
obscured due to strong absorption from the \lya\ forest.  Furthermore,
single exposures were typically too faint to fit reliable continua
even in unabsorbed regions.  Individual orders were therefore combined
by first flux calibrating each order using the response function
derived from a standard star.  The combined spectra were continuum-fit
by hand using a cubic spline.  Standards stars were also to correct
for telluric absorption.  Estimated residuals from continuum fitting
(redward of \lya) and atmospheric correction are typically within the
flux uncertainties.

\section{\oi\ Search}

\subsection{Technique}

At $z > 5$ it becomes increasingly difficult to identify individual
\lya\ lines due to the large opacity in the \lya\ forest.  Therefore,
candidate \oi\ lines must be confirmed using other ions at the same
redshift.  \siii~$\lambda 1260$ and \cii~$\lambda 1334$ are the most
useful for this purpose since their rest wavelengths place them
redward of the \lya\ forest, at least over a limited range in
redshift.  These transitions are also expected to be relatively strong
over a range in ionization conditions.  At lower redshifts where
\siii~$\lambda 1260$ has entered the forest, we are still sensitive to
\siii~$\lambda 1304$, although it is weaker than \siii~$\lambda 1260$
by a factor of 7.

We searched for \oi\ in each sightline over the redshift range between
the quasar redshift $z_{\rm QSO}$ and the redshift $z_{\rm min} = (1 +
z_{\rm QSO})\lambda_{\alpha} / \lambda_{\moi} - 1$, where \oi\ enters
the \lya\ forest.  To identify a system, we required absorption at the
same redshift in \oi\ and either \siii~$\lambda 1260$ or \cii~$\lambda
1334$, and that the line profiles for at least two of the ions be
comparable.  In practice, \siii~$\lambda 1260$ proved to be the best
indicator over the lower-redshift end of each sightline, since the
signal-to-noise ($S/N$) ratio was typically better over that region of
the spectrum.  At redshifts where \siii~$\lambda 1260$ was lost in the
forest we relied solely upon \cii\ for confirmation.  By chance, the
two \oi\ systems we identified using only \cii\ were also strong
enough to be detected in \siii~$\lambda 1304$.

We identified six \oi\ systems in total along our nine lines of sight.
Their properties are summarized in Table~2.  Four of the six systems
appear towards SDSS~J1148+5251, and we examine this potential
overabundance in detail in \S4.  Voigt profiles were fitted using the
VPFIT package written, by R. Carswell.  Due to line blending and the
modest $S/N$ ratio over some regions of the data, it was often
necessary to tie the redshift and/or Doppler parameter of one ion to
those of another whose profile was more clearly defined in order to
obtain a satisfactory fit.  In practice, the resulting total column
densities are very similar to those obtained when all parameters are
allowed to vary freely, as would be expected for systems still on the
linear part of the curve of growth.  Comments on individual systems
are given below.

\subsubsection{SDSS~J0231$-$0728: $z_{\rm sys} = 5.3364$ (Figure~1)}

This system shows a central complex in \oi, \siii, and \cii\ that is
best fit using three components with a total velocity span of $\Delta
v \approx 38$~\kms.  We measure the redshifts and $b$-parameters of
the bluest two central components from \siii.  The redshift and
$b$-parameter for the reddest central component are measured from \oi,
which is least affected by additional lines towards the red.  Leaving
all parameters untied produces a two-component fit for \cii, which is
likely due to low $S/N$.  However, the total column densities remain
virtually unchanged.

A single component at $\Delta v \approx -115$~\kms\ appears in \siii\
and \cii\ but not in \oi.  Since \oi\ is more easily ionized than
either \siii\ or \cii\ ($\Delta E_{\moi} = 13.6$~eV, $\Delta
E_{\msiii} = 16.3$~eV, $\Delta E_{\mcii} = 24.4$~eV) this component
probably arises from gas that is more highly ionized than the gas
producing the \oi\ absorption.  Three additional components appear
redward of the central \oi\ complex.  These are visible in \siii,
\cii, and \siiv\, and presumably correspond to gas in a still higher
ionization state.  We use \siii\ to constrain the redshift and
$b$-parameters for all the non-\oi\ components, although fitting
without tying any parameters produces very similar results.

\subsubsection{SDSS~J1623+3112: $z_{\rm sys} = 5.8408$ (Figure~2)}

This system shows multiple components with a total velocity span of
$\Delta v \approx 81$~\kms.  The \siii\ is strong enough that the
1304~\AA\ transition is clearly visible.  Some additional absorption
occurs along the blue edge of the \oi\ complex.  However, since no
corresponding absorption appears in \cii\ this features is unlikely to
be genuine \oi.  We use \cii\ to fix the redshifts and $b$-parameters
of the bluest two components and the reddest component for all ions.
The $z$ and $b$ values for the remaining components are tied to \oi.
The weakest \cii\ component was automatically dropped by the fitting
program.  However, this should not have a significant impact on the
summed \cii\ column density.  No \siiv\ was detected down to
$N_{\msiiv} > 10^{13}~{\rm cm^{-2}}$.

A strong line appears at a velocity separation $\Delta v \approx
-180$~\kms\ from the \oi\ complex.  Weak absorption appears at the
same velocity in \siii~$\lambda 1304$.  However, the corresponding
\cii\ region is completely covered by \mgi~$\lambda 2853$ from a
strong \mgii\ system at $z_{\rm abs} = 2.1979$.  Therefore, while
these features are plausibly associated with the \oi\ complex, their
identity cannot presently be confirmed.  We do not include these lines
when computing the relative abundances, although their \oi/\siii\ ratio
is consistent with those measured for the other \oi\ systems.

\subsubsection{SDSS~J1148+5251: $z_{\rm sys} = 6.0097$ (Figure~3)}

This system shows multiple components with a velocity span of $\Delta
v \approx 41$~\kms.  The \siii~$\lambda 1260$ transition falls in the
\lya\ forest.  However, \siii~$\lambda 1304$ is clearly visible.  The
\cii\ occurs among moderate-strength telluric absorption.  In
correcting for the atmosphere we recovered a \cii\ velocity profile
very similar to that of \oi\ and \siii.  Some additional absorption
around the \cii\ are probably telluric lines that were not completely
removed.  We achieved the best fit for \oi\ using four components
while allowing the continuum level to vary slightly.  Using fewer
components resulted in virtually the same total $N_{\moi}$, which is
to be expected if the lines lie along the linear part of the curve of
growth.  We used the $z$ and $b$ values for these four \oi\ lines to
determine the corresponding column densities for \siii\ and \cii.
Leaving all parameters untied resulted in small changes in the
$b$-parameters and in the preferred number of lines used by VPFIT (3
for \siii\ instead of 4), but negligible changes in the total column
density for each ion.  We detect no \siiv\ down to $N_{\msiiv} >
10^{12.5}~{\rm cm^{-2}}$.  This system does show \civ\ absorption in
the low-resolution NIRSPEC spectrum taken by \citet{barth03}.
However, the kinematic structure cannot be determined from their data,
and it is therefore unknown whether the \civ\ absorption arises from
the same gas that produces the lower-ionization lines.

\subsubsection{SDSS~J1148+5251: $z_{\rm sys} = 6.1293$ (Figure~4)}

The \siii~$\lambda 1260$ line for this system falls in the \lya\
forest.  However, \siii~$\lambda 1304$ is strong enough to be visible.
We tie the redshifts and $b$-parameters for \siii\ and \cii\ at $z =
6.129267$ to those for \oi, which has the cleanest profile.  Leaving
all parameters untied produces changes in the $b$-parameters that are
within the errors and negligible changes in the column densities.
Possible additional \cii\ lines appear at $\Delta v = -67$, -29, and
26 \kms\ from the strong component.  These do not appear in \oi\ and
would probably be too weak to appear in \siii~$\lambda 1304$.  We do
not include them when computing the relative abundances for this
system.  No \siiv\ was detected down to $N_{\msiiv} > 10^{12.7}~{\rm
  cm^{-2}}$.

\subsubsection{SDSS~J1148+5251: $z_{\rm sys} = 6.1968$ (Figure~5)}

The \siii~$\lambda 1260$ and \cii\ lines for this system occur in
transparent gaps within regions of mild telluric absorption.  In some
of our exposures, the \oi\ line falls on the blue wing of a strong
telluric feature.  However, the \oi\ can be clearly seen in the
uncorrected 2005 January data, where the heliocentric velocity shifted
the observed QSO spectrum further to the blue.  We tie the redshifts
and $b$-values for \oi\ and \cii\ to those of \siii, for which the
line profile is the cleanest.  Leaving all parameters untied produces
negligible changes in the column densities.

\subsubsection{SDSS~J1148+5251: $z_{\rm sys} = 6.2555$ (Figure~6)}

We use a single component for the primary fit to this system.  The
\cii\ falls on a sky emission line and hence the line profile is
rather noisy.  We therefore tied the redshift and $b$-parameter for
\cii\ to those of \siii, while $z$ and $b$ for \oi\ were allowed to
vary freely.  The fitted $z$ and $b$ values for \oi\ agree with those
for \siii\ within the fit errors.  Tying $z$ and $b$ for \oi\ to those
of \siii\ produced no significant change in $N_{\moi}$.  Both
\siii~$\lambda 1260$ and \oi\ are well fit by a single line, while the
blue edge of the \cii\ shows a possible additional component.  We fit
this component separately but do include it when computing relative
abundances.

\section{Overabundance of \oi\ systems towards SDSS~J1148+5251}

Despite the relatively small total number of detected \oi\ systems, a
noticable overabundance appears towards SDSS~J1148+5251.  It should be
noted that this object has one of the brightest continuum magnitudes
among QSOs at $z > 5$ \citep{fan03}.  We also spent considerable
integration time on this object.  Therefore, the data are among the
best in our sample.  In this section we address the null hypothesis
that the apparent imbalance in the distribution of \oi\ systems
results from a combination of incompleteness and small number
statistics.

For each sightline we used an automated scheme to determine the
absorption pathlength interval $\Delta X$ over which we are sensitive
to systems above a given \oi\ column density.  For a non-evolving
population of sources expanding with the Hubble flow, we expect to
intersect the same number of sources per unit $X$ at all redshifts,
where $X$ is defined for $\Omega = 1$ as
\begin{equation}
  \label{eq:dX}
  dX = \frac{(1+z)^2}{\sqrt{\mOm(1+z)^3 + \mOl}} dz \, .
\end{equation}
At the redshift of each pixel where we are potentially sensitive to
\oi, we inserted artificial single-component systems containing \oi,
\siii, and \cii\ with relative abundances set to the mean measured
values (see \S5) and a turbulent Doppler width of $b = 10$~\kms.  The
minimum $N_{\moi}$ at which the system could be detected by the
automated program was then determined.  A detection required that
there be significant ($4 \sigma$) absorption at the expected
wavelengths of \oi\ and at least one of the other two ions.  The
redshifts of at least two of the three fitted profiles must also agree
within a tolerance of $\Delta z_{tol} = 0.0002$, and the line widths
must agree to within a factor of two.  In the case where the fitted
profiles agree for only two of the ions, the spectrum at the
wavelength of the third ion must be consistent with the expected level
of absorption.  This allows for one of the ions to be blended with
another line or lost in \lya\ forest.  Lines that fell in the
Fraunhofer A band were not considered.  This method detected all of
the known \oi\ systems and produced only one false detection in the
real data, which was easily dismissed by visual inspection.  We note
that this method focuses on our ability to detect systems similar to
those observed towards SDSS~J1148+5251.  Multiple-component systems
such as the two at lower redshift are in general easy to identify,
even in noisy regions of the spectrum.

The results of the sensitivity measurements are shown in Figure~9.
The solid lines show the $\Delta X$ over which we are sensitive to
systems with $N_{\moi} \ge N_{\moi}^{\rm min}$.  The \oi\ column
densities for the four systems seen towards SDSS~J1148+5251 are marked
with circles.  The dashed line shows the mean $\Delta X$ as a function
of $N_{\moi}^{\rm min}$ for all nine sightlines.  While it is clear
that the path length over which we are sensitive to low-column density
systems is considerably smaller for some sightlines than for
SDSS~J1148+5251, there does exist significant coverage across the
total sample even at the $N_{\moi}$ of the weakest system in our
sample.

Completeness effects will depend on the intrinsic distribution of line
strengths, in that weaker systems will more easily be lost in
low-$S/N$ sightlines.  It is useful to define an effective absorption
pathlength interval $\Delta X^{\rm eff}$ such that, for a sightline
with a total pathlength $\Delta X^{\rm tot}$ and sensitivity $S(N)$,
we would expect to detect the same number $\mathcal{N}$ of systems
with $N \ge N_{\rm min}$ over $\Delta X^{\rm eff}$ if we had perfect
sensitivity.  For a column-density distribution $f(N) = \partial^2
\mathcal{N}/\partial N \partial X$, this gives
\begin{equation}
  \label{eq:dxeffdef}
  \Delta X^{\rm eff}\int_{N_{\rm min}}^{\infty}{f(N)\,dN} = 
     \Delta X^{\rm tot}\int_{N_{\rm min}}^{\infty}{f(N)\,S(N)\,dN}\,,
\end{equation}
where $S(N)$ is the fraction of the total pathlength over which we are
sensitive to systems with column density $N$.  The sensitivity
function can be determined directly from Figure~9.  We assume a
conventional power-law distribution in \oi\ column densities
$f(N_{\moi}) \propto N_{\moi}^{-\alpha}$.  Hence,
\begin{equation}
  \label{eq:dxeff}
  \Delta X^{\rm eff} = \Delta X^{\rm tot} \, 
                      \frac{\int_{N_{\rm min}}^{\infty}{N^{-\alpha}\,S(N)\,dN}}
                           {\int_{N_{\rm min}}^{\infty}{N^{-\alpha}\,dN}}\,.
\end{equation}
Taking into account our measured sensitivity, the maximum likelihood
estimator for $\alpha$ for all six \oi\ systems is $\hat{\alpha} =
1.7$.  A K-S test allows $1.3 \le \alpha \le 2.1$ at the 95\%
confidence level.

In Figure~10 we plot $\Delta X^{\rm eff}$ as a function of $\alpha$
taking $N_{\rm min} = 10^{13.7}$~cm$^{-2}$ as the cutoff value.  The
solid line shows $\Delta X^{\rm eff}$ for SDSS~J1148+5251, while the
dashed line shows $\Delta X^{\rm eff}$ for the remaining eight
sightlines, divided by 5.7 to match SDSS~J1148+5251 at $\alpha = 1.7$.
It is easy to see that the ratio $\Delta X^{\rm eff}_{1148} / \Delta
X^{\rm eff}_{\rm other}$ does not change significantly with $\alpha$
($\lesssim 4\%$ over $1.1 \le \alpha \le 3.0$).  This is purely a
coincidence, most likely related to the fact that the overall high
quality of the SDSS~J1148+5251 data is balanced by the placement of
the \oi\ systems in a difficult region of the spectrum.  However, it
means that the relative numbers of expected \oi\ detections for
SDSS~J1148+5251 and the remaining eight sightlines should be nearly
invariant over a wide range in plausible column density distributions.

We can now evaluate the likelihood that the imbalance of \oi\
detection results from small number sampling.  For a Poisson
distribution, the likelihood of detecting $\mathcal{N}$ systems when
the expected number of detections is $\nu$ is given by
\begin{equation}
  \label{eq:poissondef}
  P(\mathcal{N},\nu) = \frac{\nu^{\,\mathcal{N}}e^{-\nu}}{\mathcal{N}}\,.
\end{equation}
If we assume that the expected number of detections towards
SDSS~J1148+5251 is $\nu = 6\,(\Delta X^{\rm eff}_{1148} / \Delta
X^{\rm eff}_{\rm all}) \simeq 0.90$, then the probability of
detecting at least 4 systems is
\begin{equation}
  \label{eq:poisson1148}
  P(\ge 4,\nu) = \sum_{\mathcal{N} = 4}^{\infty}{P(\mathcal{N},\nu)}
                 = 1 - \sum_{\mathcal{N} = 0}^{3}{P(\mathcal{N},\nu)}
                 = 0.013 \,.
\end{equation}
Similarly, if we take the number of systems observed towards
SDSS~J1148+5251 as a representative sample and compute the expected
number of systems towards the remaining eight sightlines to be $\nu =
4\, (\Delta X^{\rm eff}_{\rm other} / \Delta X^{\rm eff}_{1148})
\simeq 23$, then the likelihood of detecting at most two systems is
\begin{equation}
  \label{eq:poissonother}
  P(\le 2,\nu) = \sum_{\mathcal{N} = 0}^{2}{P(\mathcal{N},\nu)}
               = 3.6 \times 10^{-8} \,.
\end{equation}
The probability of simultaneously detecting at least 4 systems towards
SDSS~J1148+5251 while detecting at most two systems along the
remaining eight sightlines,
\begin{equation}
  \label{eq:poissonboth}
  P \left( \mdNdX \right) = \left[ 
  1 - \sum_{\mathcal{N} = 0}^{3}{P(\mathcal{N},\mdNdX \,\Delta X^{\rm eff}_{1148})}
   \right] \, \left[
  \sum_{\mathcal{N} = 0}^{2}{P(\mathcal{N},\mdNdX \,\Delta X^{\rm eff}_{\rm other})}
   \right] \, ,
\end{equation}
has a maximum of $P(\dNdX) = 0.0018$ at $\dNdX = 0.84$, taking $\alpha
\le 3$.  Thus, assuming that the \oi\ systems are randomly
distributed, we can rule out with a high degree of confidence the
possibility that the excess number of detected systems towards
SDSS~J1148+5251 is the result of a statistical fluctuation.

\subsection{Comparison with Lower-Redshift Sightlines}

Our sample of nine $z_{\rm QSO} > 4.8$ sightlines is sufficient to
demonstrate that the apparent excess of \oi\ systems towards
SDSS~1148+5251 is highly unlikely to arise from a random spatial
distribution of absorbers.  However, the distribution may be
significantly non-random due to large-scale clustering.  For example,
if the line-of-sight towards SDSS~1148+5251 passes through an
overdense region of the IGM then we might expect to intersect more
than the typical number of absorbers.

To address this possibility, we have searched for \oi\ in the spectra
of an additional 30 QSOs with $2.6 < z_{\rm QSO} < 4.6$.  The sample
was taken entirely with HIRES for a variety of studies.  Due to the
lower QSO redshifts, the data are of significantly higher quality than
the $z_{\rm QSO} > 4.8$ sample.  All sightlines are complete down to
$N_{\moi} = 10^{13.5}~{\rm cm^{-2}}$ (apart from rare wavelength
coverage gaps) with the majority complete down to $N_{\moi} =
10^{13.0}~{\rm cm^{-2}}$.

A detailed study of the properties of lower-redshift \oi\ absorbers
will be presented elsewhere.  Here we note that 11 systems with
$N_{\moi} > 10^{13.7}~{\rm cm^{-2}}$ were identified along all 30
sightlines in the redshift intervals where \oi\ would occur redward of
the \lya\ forest.  This is comparable to but slightly larger than the
incidence rate among the eight lowest-redshift sightlines in our
$z_{\rm QSO} > 4.8$ sample.  The number density is even somewhat
higher among the $z_{\rm QSO} < 4.6$ sample since the absorption
pathlength intervals are shorter (cf.~Eq.~\ref{eq:dX}).  However, many
of the $z_{\rm QSO} < 4.6$ spectra were taken in order to study known
damped \lya\ systems (DLAs; $N_{\mhi} > 10^{20.3}~{\rm cm^{-2}}$),
which commonly exhibit \oi\ \citep[e.g.,][]{prochaska01,prochaska03}.
We would therefore expect these sightlines to contain more than the
typical number of \oi\ systems.  The important point is that no
sightline contained more than a single system.  If the excess towards
SDSS~J1148+5251 is due to the chance alignment with an overdense
region of the IGM, then such a coincidence must occur in fewer than
$\sim 2-3\%$ of all sightlines.

\subsection{Comparison with Damped \lya\ and Lyman Limit System
  Populations}

We can compare the observed number of \oi\ systems in our $z_{\rm QSO}
> 4.8$ sample with the expected number of DLAs by extrapolating from
lower-redshift DLA surveys.  \citet{s-l00} find a line number density
$n_{\rm DLA}(z) = \partial \mathcal{N}_{\rm DLA} / \partial z =
0.055\,(1+z)^{1.11}$.  This is consistent at high-$z$ with the larger
survey done by \citet{prochaska05} and with the number density
evolution \citet{rao05} obtained by combining DLA samples over $0 \le
z \le 4.5$.  From this $n_{\rm DLA}(z)$ we would expect a total of
$\sim 1.5$ DLAs total along all nine lines-of-sight in the redshift
intervals where are sensitive to \oi.  This is consistent with the two
\oi\ systems seen along the eight lower-redshift sightlines.  However,
we would expect to see only $\sim 0.2$ DLAs towards SDSS~J1148+5251,
which is inconsistent with the four we observe.

The disparity eases for SDSS~J1148+5251 if we include all Lyman limit
Systems (LLSs; $N_{\mhi} > 10^{17.2}~{\rm cm^{-2}}$).  The line number
density of LLSs at $z > 2.4$ found by \citet{peroux03}, $n_{\rm
  LLS}(z) = 0.07\,(1+z)^{2.45}$, predicts $\sim 4$ LLSs towards
SDSS~J1148+5251 in the interval where we can observe \oi.  However,
there should also be $\sim 20$ LLSs along the remaining eight
lines-of-sight.  If the excess \oi\ absorbers are associated with the
progenitors of lower-redshift LLSs then they must evolve strongly in
\oi\ at $z \gtrsim 6$.

\section{Metal Abundances}

Absorption systems at $z > 5$ provide an opportunity to study metal
enrichment when the age of the Universe was $\lesssim 1$~Gyr.  The
opacity of the \lya\ forest at these redshifts prevents us from
measuring reliable \hi\ column densities.  However, we are still able
to constrain relative abundances.  In Table~3 we summarize the Voigt
profile fitting results.  Total column densities for each system
include only components with confirmed \oi.  Relative abundances were
calculated using the solar values of \citet{grevesse98} and assuming
zero ionization corrections.  These are plotted in Figure~11.  The
error-weighted mean values for all six systems are $\langle [{\rm
  O}/{\rm Si}] \rangle = -0.04 \pm 0.06$, $\langle [{\rm C}/{\rm O}]
\rangle = -0.31 \pm 0.09$, and $\langle [{\rm C}/{\rm Si}] \rangle =
-0.34 \pm 0.07$ (2$\sigma$ errors).

The uncorrected abundances are broadly consistent with the expected
yields from Type II supernovae of low-metallicity progenitors
\citep{ww95,cl04}.  Low- and intermediate-mass stars should not
contribute significantly to the enrichment of these systems since
these stars will not have had time to evolve by $z \sim 6$
\citep[although see][]{barth03}.  We can use the O/Si ratio to
constrain the enrichment from zero-metallicity very massive stars
(VMSs, $M_{\star} \sim 140-270\,M_{\sun}$).  VMSs exploding as pair
instability supernovae are expected to yield $\sim 4$ times as much
silicon compared to Type II SNe for the same amount of oxygen
\citep{schaerer02,un02,hw02,ww95,cl04}.  We can estimate the fraction
$f^{\rm VMS}_{\rm Si}$ of silicon contributed by VMSs to these systems
using the simple mixing scheme of \citet{qw05a,qw05b},
\begin{equation}
  \label{eq:vms}
  \left( \frac{\rm O}{\rm Si} \right) = 
       \left( \frac{\rm O}{\rm Si} \right)_{\rm VMS}f_{\rm Si}^{\rm VMS} +
       \left( \frac{\rm O}{\rm Si} \right)_{\rm SN\,II}
         \left( 1 - f_{\rm Si}^{\rm VMS} \right) \, .
\end{equation}
Adopting ${\rm [O/Si]_{\rm VMS}} = -0.61$ from \citet{hw02}, and ${\rm
  [O/Si]_{\rm SN\,II}} = 0.00$ from \citet{ww95}, the mean uncorrected
O/Si gives $f^{\rm VMS}_{\rm Si} = 0.11 \pm 0.17~(2\sigma)$.  However,
this value of O/Si should be considered a lower limit.  The ionization
potential of \oi\ is smaller than that of \siii\ ($\Delta E_{\moi} =
13.6$~eV, $\Delta E_{\msiii} = 16.3$~eV).  Therefore, applying an
ionization correction would increase O/Si.  Taking $\langle [{\rm
  O}/{\rm Si}] \rangle > -0.10$ gives $f^{\rm VMS}_{\rm Si} <
0.27~(2\sigma)$.

The mean C/Si ratio is consistent with zero silicon contribution from
VMSs, although C/Si may decrease if ionization corrections are applied
($\Delta E_{\mcii} = 24.4$~eV $> \Delta E_{\msiii}$).  Dust
corrections would also decrease C/Si since silicon is more readily
depleted onto grains \citep{savage96}.  Our upper limit of $\langle
[{\rm C}/{\rm Si}] \rangle < -0.27~(2\sigma)$ therefore does not
provide an additional constraint on $f^{\rm VMS}_{\rm Si}$.  However,
the limit set by O/Si may still imply that VMSs provided a smaller
fraction of the metals in these \oi\ systems than in the general IGM.
\citet{qw05a,qw05b} have argued that VMSs at $z \gtrsim 15$ must have
contributed $\gtrsim 50\%$ of the silicon in the IGM in order to
produce ${\rm [Si/C]_{\rm IGM}} \sim 0.7$ at $z \approx 2-4$
\citep{aguirre04}.  However, the Si/C ratio in the IGM derived from
high-ionization lines depends sensitively on the choice of ionizing
backgrounds.

The high level of absorption blueward of the \lya\ emission line at $z
> 5$ prevents us from reliably fitting Voigt profiles to \hi.
However, transmission peaks in the \lya\ and \lyb\ forests near the
\oi\ redshifts allow us to set upper limits on the \hi\ column density
for two of our systems.  Fits to these systems with the strongest
allowable damping wings are shown in Figures~7 and 8, and the
corresponding limits are listed in Table~3.  The limits take into
account uncertainties in the continuum level, which was set to a
power-law $F_{\nu} \propto \lambda^{-\beta}$ with spectral index
$\beta = 0.5$, normalized at a rest wavelength of 1280\,\AA.  The
continuum uncertainty will include the error in the amplitude of the
power law at the wavelength of the damped system.  Taking $\Delta
\beta = 0.65$ to incorporate 95\% of the scatter in QSO spectral
indices \citep{richards01}, this error is $< 10\%$ for the system at
$z_{\rm sys} = 5.3364$, for which the limit on $N_{\mhi}$ was set
using \lya\ transmission, and $< 20\%$ for the system at $z_{\rm sys}
= 6.0097$, for which the limit was set using Ly$\beta$.  Larger
uncertainties in the continuum may result from errors in the relative
flux calibration and the departure from a power law.  We
conservatively estimate both of these to be $\sim 20\%$, which
translates to an uncertainty in the column density
$\sigma(\log{N_{\mhi}}) \sim 0.3$.  The upper limits on
$\log{N_{\mhi}}$ listed in Table~3 are our ``best fit'' values plus
this uncertainty. The corresponding limits on the absolute abundances
are $[{\rm O}/{\rm H}] > -2.4$ ($z_{\rm sys} = 5.3364$) and $[{\rm
  O}/{\rm H}] > -4.1$ ($z_{\rm sys} = 6.0097$).

The comoving mass density for each ion is given by
\begin{equation}
  \label{eq:omega}
  \Omega_{\rm ion} = \frac{H_0 m_{\rm ion}}{c \rho_{\rm crit}}
                    \frac{\sum{N_{\rm ion}}}{\Delta X^{\rm tot}} \, ,
\end{equation}
where $m_{\rm ion}$ is the mass of the ion, $\rho_{\rm crit} = 0.92
\times 10^{-29}~h_{70}^2~{\rm g~cm^{-3}}$ is the cosmological closure
density, and $\Delta X^{\rm tot}$ is the total pathlength interval
over which we are sensitive to \oi.  In Table~4 we list $\Omega_{\rm
  ion}$ for \oi, \siii, and \cii\ summing over all $z_{\rm QSO} > 4.8$
sightlines, over the three sightlines with $z_{\rm QSO} > 6.2$ only,
and over the sightline towards SDSS~J1148+5251 only.  The mass
densities include all confirmed absorption components, including those
without detected \oi.  However, we have made no attempt to correct for
incompleteness.  The densities of low-ionization metal species along
our $z_{\rm QSO} > 6.2$ sightlines are comparable to the densities of
the corresponding high-ionization species at lower-redshift.
\citet{songaila05} measured $\Omega(\mbox{\civ}) \sim 3 \times
10^{-8}$ and $\Omega(\mbox{\siiv}) \sim 1 \times 10^{-8}$ over $1.5 <
z < 4.5$, with a drop in $\Omega(\mbox{\civ})$ by at least a factor of
two at $z > 5$.  A decrease in high-ionization metals and a correlated
rise in their low-ionization counterparts with redshift might signal a
change in the ionization state of the $z > 5$ IGM.  However, coverage
of \civ\ does not currently extend to $z \sim 6$, and our \oi\ sample
is probably too small to make a meaningful comparison.

\section{Discussion}

Although we observe an overabundance of \oi\ at $z > 6$ along one of
our highest-redshift sightlines, it is not clear that these four
systems constitute a ``forest'' which would signal a predominantly
neutral IGM.  In the reionization phase prior to the overlap of
ionized bubbles, \citet{oh02} predicts tens to hundreds of \oi\ lines
with $N_{\moi} > 10^{14}~{\rm cm^{-2}}$, although the number may be
lower if the enriched regions have very low metallicity ($Z <
10^{-2.5}Z_{\sun}$), or if the volume filling factor of metals is
small ($f_{Z} \ll 0.01$).  The number of \oi\ systems we detect is
more consistent with his predictions for the post-overlap phase,
although the expected number of lines depends on the ionizing
efficiency of the metal-producing galaxies.  The \citet{oh02} model
also predicts lines with $N_{\moi} > 10^{15}~{\rm cm^{-2}}$, which we
do not observe.

The significance of the excess \oi\ towards SDSS~J1148+5251 is
complicated by the presence of \lya\ and \lyb\ transmission along that
sightline.  The \oi\ systems occur interspersed in redshift with the
\lyb\ peaks, with the velocity separation from the nearest peak
ranging from $\Delta v \sim 500$ to $3000$~\kms\ ($\Delta l \sim 5$ to
$30$~comoving Mpc).  The \oi\ absorbers also do not appear to be
clustered.  Along a total redshift interval $\Delta z_{\rm tot} =
0.49$ ($\sim 200$~comoving Mpc) where we are sensitive to \oi, the
detected systems are separated by $\Delta z = 0.06$ to $0.12$ ($\Delta
l = 24$ to $50$~comoving Mpc).  If any significantly neutral regions
are present along this sightline, then they occur within or near
regions that are highly ionized.  In contrast, we find no \oi\ towards
SDSS~J1030+0524, despite the fact that its spectrum shows complete
absorption in \lya\ and \lyb\ over $\Delta z \sim 0.2$.  The
SDSS~J1030+0524 sightline may pass through a region of the IGM that is
still largely neutral and not yet enriched.  Alternatively, there may
be no \oi\ because the sightline is highly ionized.  In that case, a
large-scale density enhancement could still produce the high \lya\ and
\lyb\ optical depth \citep{wyithe05}.

The overabundance of \oi\ together with the transmission in \lya\ and
\lyb\ indicates that the SDSS~J1148+5251 sightline has experienced
both significant enrichment \textit{and} ionization.  One explanation
is that we are looking along a filament.  The total comoving distance
between \oi\ systems towards SDSS~J1148+5251 is $\approx 101$~Mpc,
which is smaller than the largest known filamentary structures at low
redshift \citep[e.g.,][]{gott05}.  If the line-of-sight towards
SDSS~1148+5251 runs along a filament at $z \sim 6$ then we might
expect to intersect more than the average number of absorbers.
Galaxies along such a structure might produce enough ionizing
radiation to create the transmission gaps observed in the \lya\ and
\lyb\ forests while still allowing pockets of neutral material to
persist if the ionizing background is sufficiently low.  If we assume
that every $z \sim 6$ galaxy is surrounded by an enriched halo, then
we can calculate the typical halo radius
\begin{eqnarray}
  \label{eq:rhalo}
   R_{\rm halo} & \sim & \left[ \frac{\mathcal N}{\pi (1+z)^2 
           \left<\Phi_{\rm gal}\right> \delta_{\rm gal} \Delta D} \right]^{1/2}\\
   \nonumber & \sim & 35~{\rm proper~kpc} \, 
            \left(\frac{\mathcal N}{4}\right)^{1/2}
            \left(\frac{1+z}{7.2}\right)^{-1}
            \left(\frac{\left<\Phi_{\rm gal}\right>}
                        {0.01\,{\rm Mpc^{-3}}}\right)^{-1/2}  \\
    & &     \times \left(\frac{\delta_{\rm gal}}{10}\right)^{-1/2}
            \left(\frac{\Delta D}{200\,{\rm Mpc}}\right)^{-1/2} \, .
\end{eqnarray}
Here, $\Delta D$ is the total comoving distance over which we are
sensitive to \oi.  We have used the mean comoving number of galaxies
$\left<\Phi_{\rm gal}\right>$ at $z \sim 6$ measured down to the limit
of the Hubble Ultra Deep Field \citep[UDF;][]{yan04,bouwens05}, and
assumed an overdensity of galaxies $\delta_{\rm gal} = \Phi_{\rm gal}
/ \left<\Phi_{\rm gal}\right> = 10$ along this sightline.  The
absorption could be due to filled halos, shells, or other structures
surrounding the galaxies.  Note that by inverting Eq.~\ref{eq:rhalo}
we can rule out the possibility that these \oi\ systems arise from the
central star-forming regions of galaxies similar those observed in the
UDF.  The small size $R_{\rm half-light} \sim 0.8$~kpc of galaxies at
$z \sim 6$ \citep{bouwens05} would require either a local galaxy
overdensity $\delta_{\rm gal} \sim 10^4$ or a large number of faint
galaxies to produce the observed absorption systems.

At present we can place relatively few physical constraints on these
absorbers.  Assuming charge-exchange equilibrium
(Eq.~\ref{eq:equilib}), we can estimate the \hi\ column densities as a
function of metallicity,
\begin{equation}
  \label{eq:NHI}
  \log{N_{\mhi}} =  16.9 + 
     \log{\left(\frac{N_{\moi}}{10^{13.7}~{\rm cm^{-2}}}\right)} - 
       \rm{[O/H]}\,.
\end{equation}
For solar metallicities, even the weakest \oi\ system in our sample
must be nearly optically thick.  Metal-poor absorbers with [O/H] $<
-2$ would require column densities $N_{\mhi} \gtrsim 10^{19}~{\rm
  cm^{-2}}$, which is consistent with the indication from the relative
metal abundances that these absorbers are largely neutral.  

Finally, the excess \oi\ absorbers are distinguished by their small
velocity dispersions.  Three out of the four systems towards
SDSS~J1148+5251 have single components with Doppler widths $b \approx
5-8$~\kms.  We may be seeing only the low-ionization component of a
kinematically more complex structure such as a galactic outflow
\citep[e.g.,][]{simcoe05}.  However, if there are additional
high-ionization components that would appear in \civ\ and \siiv\ then
we might expect to see components in \cii\ and possibly \siii\ without
\oi.  This is the case in the $z_{\rm sys} = 5.3364$ system towards
SDSS~J0231-0728.  The $z_{\rm sys} = 6.1293$ and $z_{\rm sys} =
6.2550$ systems towards SDSS~J1148+5251 show potential \cii\ lines
that do not appear in \oi, but these are relatively weak, and in the
$z_{\rm sys} = 6.1293$ system there is no apparent strong \siiv.
Among the 11 \oi\ absorbers in the $z < 4.6$ sample, only three have a
single \oi\ component with a velocity width comparable to the three
narrowest systems towards SDSS~J1148+5251.  The fact that the excess
\oi\ systems appear kinematically quiescent further suggests that they
are a distinct class of absorbers.

\section{Summary}

We have conducted a search for \oi\ in the high-resolution spectra of
9 QSOs at $4.8 \le z_{\rm QSO} \le 6.4$.  In total, we detect six
systems in the redshift intervals where \oi~$\lambda 1302$ falls
redward of the \lya\ forest.  Four of these lie towards
SDSS~J1148+5251 ($z_{\rm QSO} = 6.42$).  This imbalance is not easily
explained by small number statics or by varying sensitivity among the
data.  A search at lower redshift revealed no more than than one \oi\
system per sightline in a high-quality sample of 30 QSOs with $2.6 \le
z_{\rm QSO} \le 4.6$.  The number of excess systems is significantly
larger than the expected number of damped \lya\ systems at $z \sim 6$,
but similar to the predicted number of Lyman limit systems.

The six systems have mean relative abundances $\langle [{\rm O}/{\rm
  Si}] \rangle = -0.04 \pm 0.06$, $\langle [{\rm C}/{\rm O}] \rangle =
-0.31 \pm 0.09$, and $\langle [{\rm C}/{\rm Si}] \rangle = -0.34 \pm
0.07$ ($2\sigma$ errors), assuming no ionization corrections.  These
abundances are consistent with enrichment dominated by Type II
supernovae.  The lower limit on the oxygen to silicon ratio $\langle
[{\rm O}/{\rm Si}] \rangle > -0.10$ is insensitive to ionization
corrections and limits the contribution of silicon from VMSs to $<
30\%$ in these systems.  The upper limits on the ratio of carbon to
oxygen ${\rm [C/O]} \le -0.13$, and carbon to silicon ${\rm [C/Si]}
\le -0.20$ are also insensitive to ionization.  Integrating over only
the confirmed absorption lines, the mean ionic comoving mass densities
along the three $z_{\rm QSO} > 6.2$ sightlines are $\Omega_{\moi}
\approx 7.0 \times 10^{-8}$, $\Omega_{\msiii} \approx 9.6 \times
10^{-9}$, and $\Omega_{\mcii} \approx 1.5 \times 10^{-8}$.

The excess \oi\ systems towards SDSS~J1148+5251 occur in redshift
among \lya\ and \lyb\ transmission features, which indicates that the
IGM along this sightline has experienced both enrichment \textit{and}
significant ionization.  The sightline may pass along a filament,
where galaxies are producing enough ionizing photons create the
transmission gaps while still allowing a few neutral absorbers to
persist.  In contrast, no \oi\ was observed towards SDSS~J1030+0524,
whose spectrum shows complete Gunn-Peterson absorption.  The
SDSS~J1030+0524 sightline may pass through an underdense region
containing very few sources that would either enrich or ionize the
IGM.  Alternatively, the sightline may be highly ionized, with a
large-scale density enhancement producing strong line-blending in the
\lya\ and \lyb\ forests.  Future deep imaging of these fields and
near-infrared spectroscopy covering high-ionization metal lines will
allow us to distinguish between these scenarios and clarify the
nature of the excess \oi\ absorbers.

\acknowledgments

The authors would like to thank the anonymous referee for helpful
comments, Peng Oh for sharing his recent results on charge-exchange
equilibrium ahead of publication, and Bob Carswell for VPFIT.  WLWS
gratefully acknowledge support from the NSF through grants AST
99-00733 and AST 02-06067.  MR has been supported by the NSF under
grant AST 00-98492.  RAS has been supported by the MIT Pappalardo
Fellowship program.  Finally, we thank the Hawaiian people for the
opportunity to conduct observations from Mauna Kea.  Without their
hospitality this work would not have been possible.  In particular, we
would like to thank Lono, who granted us clear skies during the
overcast months of 2005 January and February.

\clearpage
   
\begin{deluxetable}{lccc}
   \tabletypesize{\scriptsize}
   \tablewidth{0pt}
   \centering
   \tablecaption{Summary of Observations}
   \tablehead{
        \colhead{QSO} & \colhead{$z_{\rm QSO}$} & 
        \colhead{Dates} & \colhead{$\Delta t_{\exp}$~(hrs)\tablenotemark{a}}
   }
   \startdata
   SDSS~J2225$-$0014 & 4.87 & 2004 Jun & 6.7\tablenotemark{b} \\
   SDSS~J1204$-$0021 & 5.09 & 2005 Jan - Feb & 6.7 \\
   SDSS~J0915$+$4244 & 5.20 & 2005 Jan - Feb & 10.8 \\
   SDSS~J0231$-$0728 & 5.42 & 2005 Jan - Feb & 10.0 \\
   SDSS~J0836$+$0054 & 5.80 & 2003 Feb - 2005 Jan & 21.7\tablenotemark{c} \\
   SDSS~J0002$+$2550 & 5.82 & 2005 Jan - Jun & 4.2 \\
   SDSS~J1623$+$3112 & 6.22 & 2005 Jun & 12.5 \\
   SDSS~J1030$+$0524 & 6.30 & 2005 Feb & 12.0 \\
   SDSS~J1148$+$5251 & 6.42 & 2003 Feb - 2005 Feb & 22.0\tablenotemark{d} \\
   \enddata
   \tablenotetext{a}{All exposures were taken with the upgraded HIRES detector
                     unless otherwise noted.}
   \tablenotetext{b}{Taken with the old HIRES detector}
   \tablenotetext{c}{9.2 hrs taken with the old HIRES detector}
   \tablenotetext{d}{4.2 hrs taken with the old HIRES detector}
\end{deluxetable}

\clearpage

\begin{deluxetable}{llll}
   \tabletypesize{\scriptsize}
   \tablewidth{0pt}
   \centering
   \tablecaption{Measured Properties of \oi\ Systems}
   \tablehead{ \colhead{Ion} & \colhead{$z$} & \colhead{$b$~(\kms)} & 
               \colhead{$\log{N}~{\rm (cm^{-2})}$}
   }
   \startdata
   \cutinhead{\textit{SDSS~J0231$-$0728:} $z_{\rm sys} = 5.3364$}
   \siii & $5.333982 \pm 0.000008$ & $ 5.38 \pm 0.58$ & $12.786 \pm 0.055$ \\
   \cii\tablenotemark{a}   & $5.333982$ & $ 5.38$ & $13.584 \pm 0.068$ \\
   \hline
   \siii & $5.335928 \pm 0.000020$ & $ 6.91 \pm 1.33$ & $12.514 \pm 0.079$ \\
   \siii & $5.336337 \pm 0.000019$ & $ 7.62 \pm 1.60$ & $12.978 \pm 0.087$ \\
   \siii\tablenotemark{b} & $5.336721$ & $12.96$ & $12.711 \pm 0.135$ \\
   \oi\tablenotemark{a}    & $5.335928$ & $ 6.91$ & $13.982 \pm 0.086$ \\
   \oi\tablenotemark{a}    & $5.336337$ & $ 7.62$ & $14.138 \pm 0.129$ \\
   \oi   & $5.336721 \pm 0.000072$ & $12.96 \pm 3.23$ & $13.972 \pm 0.143$ \\
   \cii\tablenotemark{a}   & $5.335928$ & $ 6.91$ & $13.298 \pm 0.077$ \\
   \cii\tablenotemark{a}   & $5.336337$ & $ 7.62$ & $13.454 \pm 0.119$ \\
   \cii\tablenotemark{b}  & $5.336721$ & $12.96$ & $13.314 \pm 0.138$ \\
   \hline
   \siii & $5.337841 \pm 0.000014$ & $10.89 \pm 0.98$ & $12.678 \pm 0.040$ \\
   \siii & $5.338474 \pm 0.000043$ & $11.71 \pm 3.39$ & $12.233 \pm 0.098$ \\
   \siii & $5.339301 \pm 0.000016$ & $10.44 \pm 1.13$ & $12.316 \pm 0.063$ \\
   \cii\tablenotemark{a}   & $5.337841$ & $10.89$ & $13.512 \pm 0.051$ \\
   \cii\tablenotemark{a}   & $5.338474$ & $11.71$ & $12.880 \pm 0.161$ \\
   \cii\tablenotemark{a}   & $5.339301$ & $10.45$ & $13.074 \pm 0.097$ \\
   \siiv\tablenotemark{a}  & $5.337841$ & $10.89$ & $12.848 \pm 0.051$ \\
   \siiv\tablenotemark{a}  & $5.338474$ & $11.71$ & $12.142 \pm 0.192$ \\
   \siiv\tablenotemark{a}  & $5.339301$ & $10.44$ & $12.950 \pm 0.049$ \\
   \cutinhead{\textit{SDSS~J1623+3112:} $z_{\rm sys} = 5.8408$}
   \oi\tablenotemark{c}   & $5.836481 \pm 0.000012$ & $ 5.01  \pm 0.59$ & $14.376 \pm 0.075$ \\
   \oi\tablenotemark{c}   & $5.836806 \pm 0.000017$ & $ 4.44  \pm 1.57$ & $14.216 \pm 0.111$ \\
   \siii\tablenotemark{c} & $5.836423 \pm 0.000029$ & $ 3.63  \pm 2.66$ & $12.746 \pm 0.182$ \\
   \siii\tablenotemark{c} & $5.836799 \pm 0.000060$ & $10.82  \pm 4.29$ & $12.985 \pm 0.124$ \\
   \hline
   \siii\tablenotemark{d} & $5.839938$ & $ 6.37$ & $13.279 \pm 0.182$ \\
   \siii\tablenotemark{d} & $5.840138$ & $ 4.13$ & $13.063 \pm 0.340$ \\
   \siii\tablenotemark{b} & $5.840456$ & $ 6.92$ & $12.931 \pm 0.196$ \\
   \siii\tablenotemark{b} & $5.840798$ & $ 5.32$ & $13.488 \pm 0.060$ \\
   \siii\tablenotemark{b} & $5.841113$ & $ 7.33$ & $13.271 \pm 0.078$ \\
   \siii\tablenotemark{b} & $5.841456$ & $ 1.77$ & $11.655 \pm 1.018$ \\
   \siii\tablenotemark{d} & $5.841780$ & $ 6.23$ & $12.854 \pm 0.070$ \\
   \oi\tablenotemark{d}   & $5.839938$ & $ 6.37$ & $14.214 \pm 0.212$ \\
   \oi\tablenotemark{d}   & $5.840138$ & $ 4.13$ & $14.152 \pm 0.242$ \\
   \oi   & $5.840456 \pm 0.000032$ & $ 6.92 \pm 4.21$ & $13.734 \pm 0.206$ \\
   \oi   & $5.840798 \pm 0.000014$ & $ 5.32 \pm 0.94$ & $14.384 \pm 0.084$ \\
   \oi   & $5.841113 \pm 0.000025$ & $ 7.33 \pm 1.48$ & $14.316 \pm 0.073$ \\
   \oi   & $5.841456 \pm 0.000014$ & $ 1.77 \pm 1.72$ & $13.518 \pm 0.150$ \\
   \oi\tablenotemark{d}   & $5.841780$ & $ 6.23$ & $13.793 \pm 0.027$ \\
   \cii  & $5.839938 \pm 0.000055$ & $ 6.37 \pm 1.56$ & $13.811 \pm 0.231$ \\
   \cii  & $5.840138 \pm 0.000039$ & $ 4.13 \pm 2.50$ & $13.955 \pm 0.249$ \\
   \cii\tablenotemark{b}  & $5.840456$ & $ 6.92$ & $13.431 \pm 0.193$ \\
   \cii\tablenotemark{b}  & $5.840798$ & $ 5.32$ & $13.771 \pm 0.093$ \\
   \cii\tablenotemark{b}  & $5.841113$ & $ 7.33$ & $13.316 \pm 0.101$ \\
   \cii  & $5.841780 \pm 0.000007$ & $ 6.23 \pm 0.51$ & $13.736 \pm 0.055$ \\
   \cutinhead{\textit{SDSS~J1148+5251:} $z_{\rm sys} = 6.0097$}
   \siii\tablenotemark{b} & $6.009119$ & $ 7.02$ & $12.126 \pm 0.592$ \\
   \siii\tablenotemark{b} & $6.009290$ & $ 3.36$ & $12.770 \pm 0.137$ \\
   \siii\tablenotemark{b} & $6.009717$ & $ 8.44$ & $13.323 \pm 0.037$ \\
   \siii\tablenotemark{b} & $6.010070$ & $ 4.94$ & $12.331 \pm 0.261$ \\
   \oi   & $6.009119 \pm 0.000112$ & $ 7.02 \pm 3.18$ & $13.593 \pm 0.371$ \\
   \oi   & $6.009290 \pm 0.000012$ & $ 3.36 \pm 1.28$ & $14.071 \pm 0.113$ \\
   \oi   & $6.009717 \pm 0.000010$ & $ 8.44 \pm 0.96$ & $14.487 \pm 0.030$ \\
   \oi   & $6.010070 \pm 0.000034$ & $ 4.94 \pm 1.60$ & $13.516 \pm 0.178$ \\
   \cii\tablenotemark{b}  & $6.009119$ & $ 7.02$ & $13.180 \pm 0.328$ \\
   \cii\tablenotemark{b}  & $6.009290$ & $ 3.36$ & $13.281 \pm 0.286$ \\
   \cii\tablenotemark{b}  & $6.009717$ & $ 8.44$ & $14.013 \pm 0.057$ \\
   \cii\tablenotemark{b}  & $6.010070$ & $ 4.94$ & $13.044 \pm 0.182$ \\
   \cutinhead{\textit{SDSS~J1148+5251:} $z_{\rm sys} = 6.1293$}
   \cii\tablenotemark{c}  & $6.127673 \pm 0.000039$ & $ 6.11 \pm 3.03$ & $12.916 \pm 0.118$ \\
   \cii\tablenotemark{c}  & $6.128589 \pm 0.000037$ & $ 8.08 \pm 2.47$ & $13.101 \pm 0.092$ \\
   \hline
    \siii\tablenotemark{b} & $6.129267$ & $ 6.51$ & $13.273 \pm 0.050$ \\
   \oi   & $6.129267 \pm 0.000006$ & $ 6.51 \pm 0.49$ & $14.606 \pm 0.096$ \\
   \cii\tablenotemark{b}  & $6.129267$ & $ 6.51$ & $13.744 \pm 0.070$ \\
   \hline
   \cii\tablenotemark{c}  & $6.129885 \pm 0.000017$ & $ 2.21 \pm 1.80$ & $13.161 \pm 0.186$ \\
   \cutinhead{\textit{SDSS~J1148+5251:} $z_{\rm sys} = 6.1968$}
   \siii & $6.196787 \pm 0.000014$ & $ 7.82 \pm 0.84$ & $12.277 \pm 0.044$ \\
   \oi\tablenotemark{a}    & $6.196787$ & $ 7.82$ & $13.727 \pm 0.061$ \\
   \cii\tablenotemark{a}   & $6.196787$ & $ 7.82$ & $13.027 \pm 0.072$ \\
   \cutinhead{\textit{SDSS~J1148+5251:} $z_{\rm sys} = 6.2555$}
   \cii\tablenotemark{c}  & $6.255234 \pm 0.000131$ & $12.13 \pm 5.51$ & $13.159 \pm 0.229$ \\
   \hline
   \siii & $6.255496 \pm 0.000004$ & $ 4.30 \pm 0.27$ & $12.900 \pm 0.040$ \\
   \oi   & $6.255465 \pm 0.000015$ & $ 5.08 \pm 1.04$ & $14.075 \pm 0.087$ \\
   \cii\tablenotemark{a}   & $6.255496$ & $ 4.30$ & $13.596 \pm 0.114$ \\
   \enddata
   \tablenotetext{a}{The $z$- and $b$-values for this component were tied
      to those for \siii.}
   \tablenotetext{b}{The $z$- and $b$-values for this component were tied
      to those for \oi.}
   \tablenotetext{c}{Unconfirmed component}
   \tablenotetext{d}{The $z$- and $b$-values for this component were tied
      to those for \cii.}
\end{deluxetable}

\clearpage

\begin{deluxetable}{lcccccccc}
   \tabletypesize{\scriptsize}
   \centering
   \tablecaption{Abundance Measurements}
   \tablewidth{0pt}
   \tablehead{ 
        \colhead{Sightline} & \colhead{$z_{\rm sys}$} & 
        \colhead{$\log{N_{\moi}^{\rm tot}}$\tablenotemark{a}} &
        \colhead{$\log{N_{\msiii}^{\rm tot}}$\tablenotemark{a}} & 
        \colhead{$\log{N_{\mcii}^{\rm tot}}$\tablenotemark{a}} & 
        \colhead{[O/Si]\tablenotemark{b}} & 
        \colhead{[C/O]\tablenotemark{b}} & 
        \colhead{[C/Si]\tablenotemark{b}} & 
        \colhead{$\log{N_{\mhi}}$}
   }
   \startdata 
   SDSS~J1148$+$5251 & 6.2555 & $14.08 \pm 0.09$ & $12.90 \pm 0.04$ & $13.60 \pm 0.11$ & $-0.11 \pm 0.08$ & $-0.17 \pm 0.15$ & $-0.27 \pm 0.12$ &  \nodata  \\ % [O/H] > -3.6
   SDSS~J1148$+$5251 & 6.1968 & $13.73 \pm 0.06$ & $12.28 \pm 0.04$ & $13.03 \pm 0.07$ & $ 0.17 \pm 0.07$ & $-0.39 \pm 0.10$ & $-0.22 \pm 0.08$ &  \nodata  \\
   SDSS~J1148$+$5251 & 6.1293 & $14.61 \pm 0.10$ & $13.27 \pm 0.05$ & $13.74 \pm 0.07$ & $ 0.05 \pm 0.10$ & $-0.55 \pm 0.13$ & $-0.50 \pm 0.09$ &  \nodata  \\
   SDSS~J1148$+$5251 & 6.0097 & $14.70 \pm 0.05$ & $13.48 \pm 0.05$ & $14.17 \pm 0.07$ & $-0.07 \pm 0.06$ & $-0.22 \pm 0.09$ & $-0.28 \pm 0.08$ &  $< 21.9$\tablenotemark{c} \\ % [O/H] > -4.1
   SDSS~J1623$+$3112 & 5.8408 & $14.96 \pm 0.06$ & $13.98 \pm 0.06$ & $14.50 \pm 0.09$ & $-0.31 \pm 0.09$ & $-0.15 \pm 0.13$ & $-0.45 \pm 0.11$ &  \nodata  \\
   SDSS~J0231$-$0728 & 5.3364 & $14.52 \pm 0.07$ & $13.25 \pm 0.06$ & $13.84 \pm 0.07$ & $-0.02 \pm 0.07$ & $-0.37 \pm 0.12$ & $-0.39 \pm 0.09$ &  $< 20.1$\tablenotemark{c} \\ % [O/H] > -2.4
   \enddata
   \tablenotetext{a}{Total ionic column densities summed over all
     components with confirmed \oi\ absorption}
   \tablenotetext{b}{Relative abundances using the solar values of
     \citet{grevesse98} and assuming no ionization corrections}
   \tablenotetext{c}{Upper limit including uncertainty in the continuum level}
\end{deluxetable}

\clearpage

\begin{deluxetable}{lccc}
   \tabletypesize{\scriptsize}
   \centering
   \tablecaption{Comoving Mass Densities\tablenotemark{a}}
   \tablewidth{0pt}
   \tablehead{ 
        \colhead{Sightlines} & \colhead{$\Omega(\moi)$} & 
        \colhead{$\Omega(\msiii)$} & \colhead{$\Omega(\mcii)$} \\
        & \colhead{($\times 10^{-8}$)} & \colhead{($\times 10^{-9}$)} &
        \colhead{($\times 10^{-8}$)}
   }
   \startdata 
   All                 & $ 3.2 \pm 0.2$ & $4.5 \pm 0.4$ & $0.8 \pm 0.1$ \\
   $z_{\rm QSO} > 6.2$ & $ 7.0 \pm 0.6$ & $9.6 \pm 0.9$ & $1.5 \pm 0.2$ \\
   SDSS~J1148+5251     & $10.2 \pm 1.0$ & $9.8 \pm 0.7$ & $1.8 \pm 0.2$ \\
   \enddata

   \tablenotetext{a}{For each ion, the value of $\Omega_{\rm ion}$
     computed from Eq.~\ref{eq:omega} includes all \textit{confirmed}
     components listed in Table~2 along the indicated lines of sight,
     including those components without corresponding \oi.  These
     values are lower limits in the sense that we have made no attempt
     to correct for incompleteness.  Errors are 1$\sigma$ and reflect
     only the uncertainty in the summed column densities.}

\end{deluxetable}

\clearpage

\begin{figure}
  \epsscale{0.6}
  \centering
  \plotone{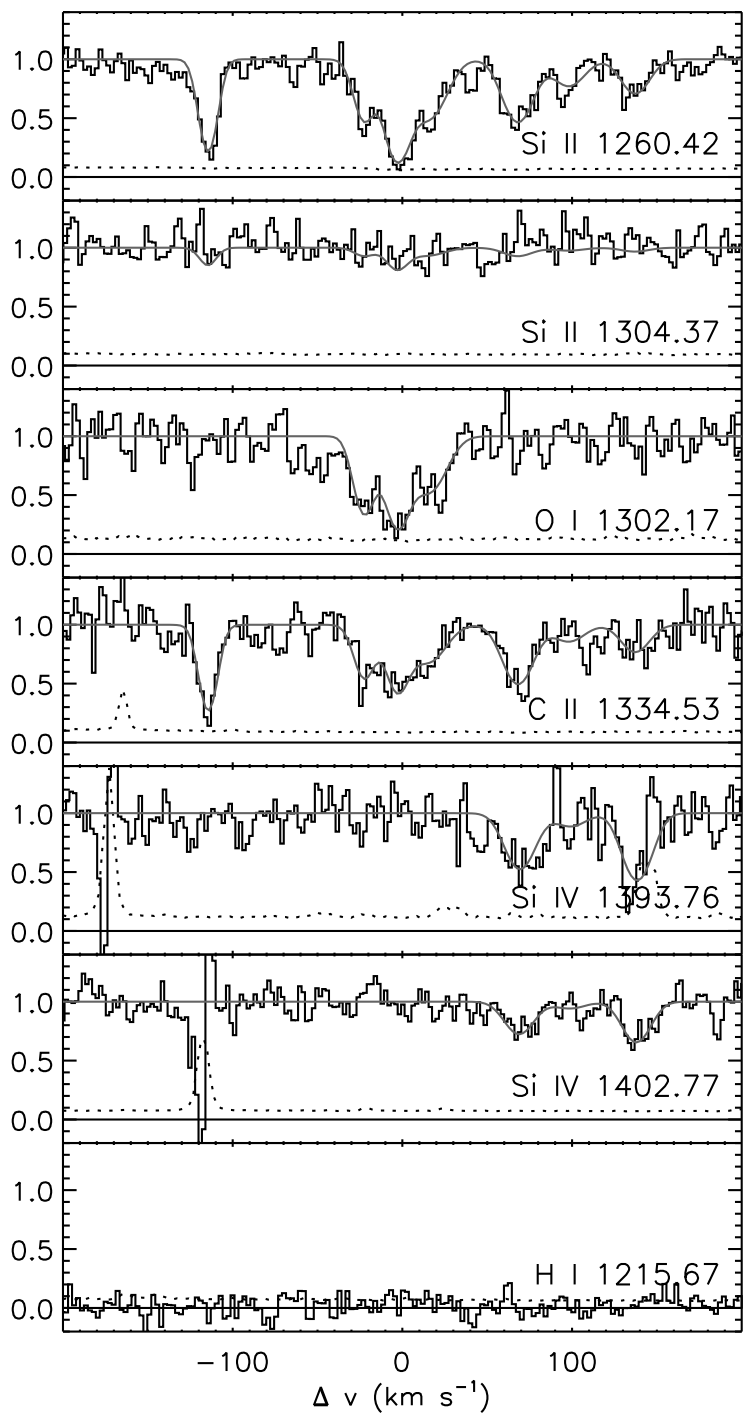}
  \figcaption{Absorption lines for the $z_{\rm sys}=5.3364$ \oi\
    system towards SDSS~J0231$-$0728.  Regions of the spectrum covering
    individual transitions have been shifted onto a common velocity
    scale.  The \textit{histogram} shows the normalized flux.  The
    \textit{solid line} shows the best Voigt profile fit.  The
    \textit{dotted line} shows the $1\sigma$ flux error.}
  \label{oi_z5.3364}
\end{figure}

\clearpage

\begin{figure}
  \epsscale{0.6}
  \centering
  \plotone{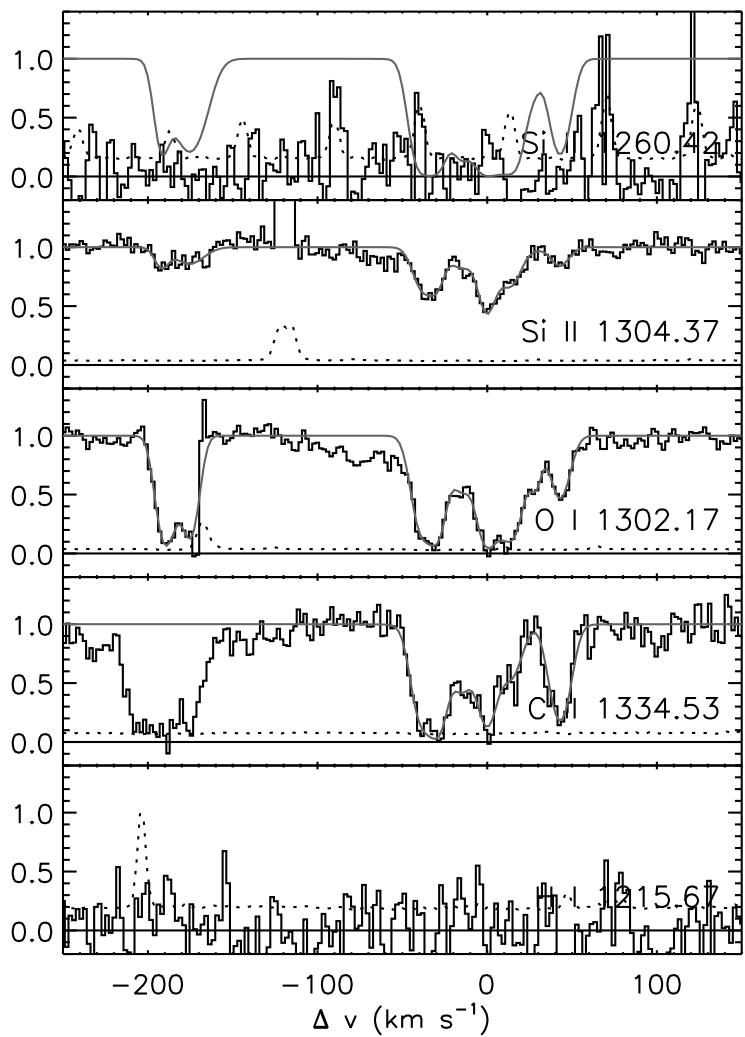}

  \figcaption{Absorption lines for the $z_{\rm sys}=5.8408$ \oi\
    system towards SDSS~1623+3112.  See Figure~1 for details.  The
    sharp features around \siii~$\lambda 1260$ are noise in the \lya\
    forest.  The components at $\Delta v \approx -180$~\kms\ are
    unconfirmed due to the strong $z_{\rm abs} = 2.1979$ \mgi\
    absorption at the wavelength where we would expect to see \cii.}

  \label{oi_z5.8408}
\end{figure}

\clearpage

\begin{figure}
  \epsscale{0.6}
  \centering
  \plotone{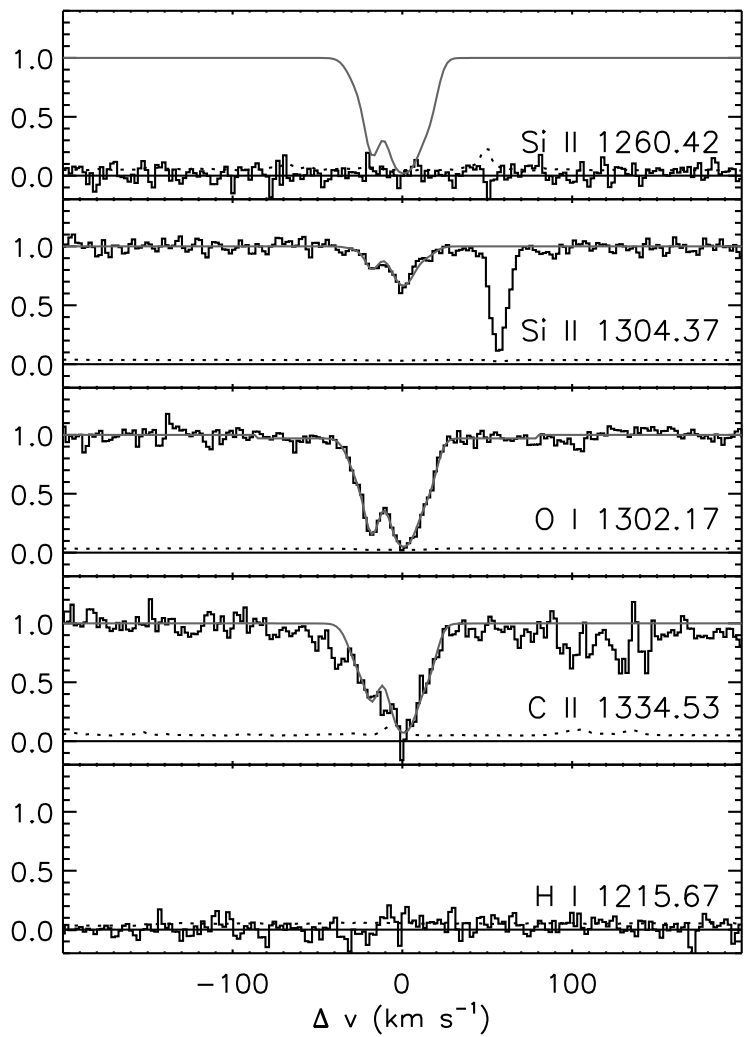}
  \figcaption{Absorption lines for the $z_{\rm sys}=6.0097$ \oi\
    system towards SDSS~1148+5251.  See Figure~1 for details.
    \siii~$\lambda 1260$ falls in the \lya\ forest.  The strong line
    at $\Delta v = 57$~\kms\ from \siii~$\lambda 1304$ is
    \siii~$\lambda 1260$ at $z = 6.2555$.  The absorption features at
    $\Delta v \approx -40$~\kms\ and $\Delta v \approx 90-140$~\kms\
    from \cii\ are probably telluric absorption that has not been
    fully removed. }
  \label{oi_z6.0097}
\end{figure}

\clearpage

\begin{figure}
  \epsscale{0.6}
  \centering
  \plotone{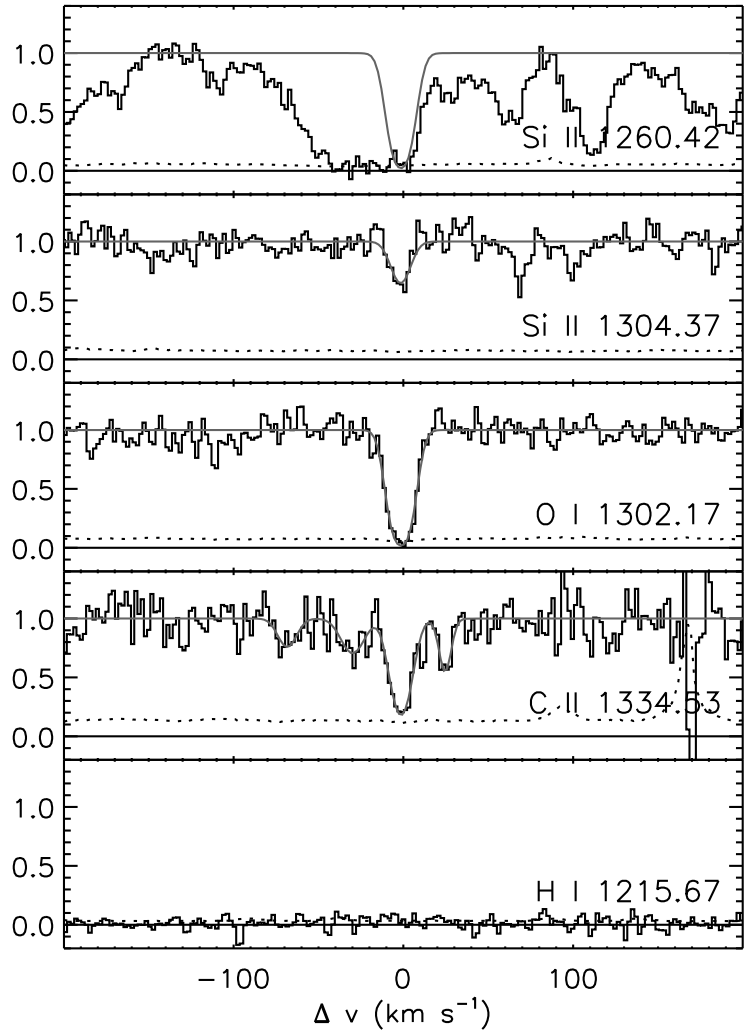}
  \figcaption{Absorption lines for the $z_{\rm sys}=6.1293$ \oi\
    system towards SDSS~1148+5251.  See Figure~1 for details.  The
    features around \siii~$\lambda 1260$ are \lya\ absorption in
    the quasar proximity region.  The \cii\ components at
    $\Delta v = -67, -29$ and 26~\kms\ are unconfirmed.}
  \label{oi_z6.1293}
\end{figure}

\clearpage

\begin{figure}
  \epsscale{0.6}
  \centering
  \plotone{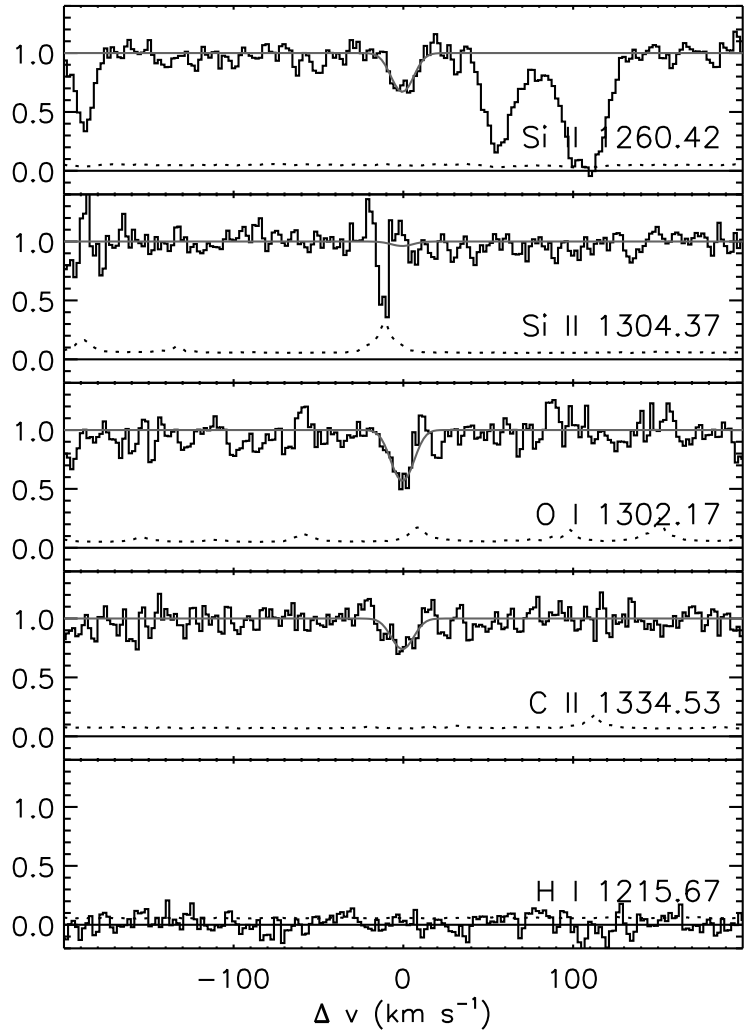}
  \figcaption{Absorption lines for the $z_{\rm sys}=6.1968$ \oi\
    system towards SDSS~1148+5251.  See Figure~1 for details.  The
    features near \siii~$\lambda 1260$ are unrelated metal lines.  The
    sharp feature near \siii~$\lambda 1304$ is the residual from a
    strong skyline.}
  \label{oi_z6.1968}
\end{figure}

\clearpage

\begin{figure}
  \epsscale{0.6}
  \centering
  \plotone{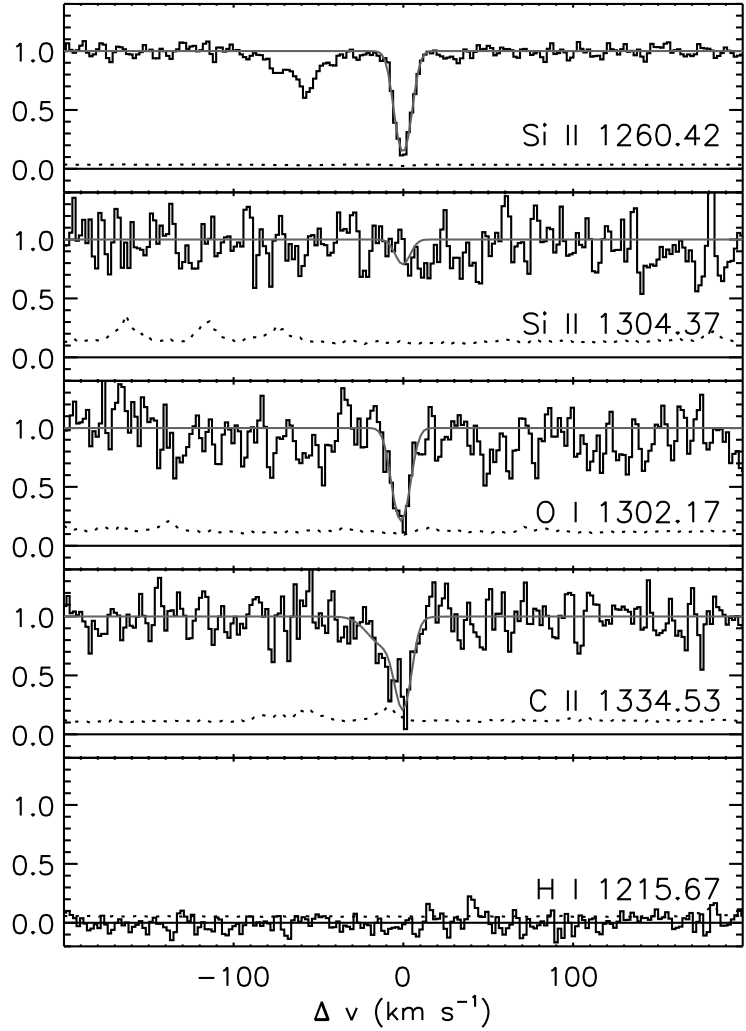}

  \figcaption{Absorption lines for the $z_{\rm sys}=6.2555$ \oi\
    system towards SDSS~1148+5251.  See Figure~1 for details.  The
    absorption at $\Delta v \approx -60$~\kms\ from \siii~$\lambda
    1260$ is \siii~$\lambda 1304$ at $z = 6.0097$.  The \cii\ fit includes 
    an unconfirmed component on the blue edge of the profile.}

  \label{oi_z6.2555}
\end{figure}

\clearpage

\begin{figure}
  \epsscale{0.75}
  \centering
  \plotone{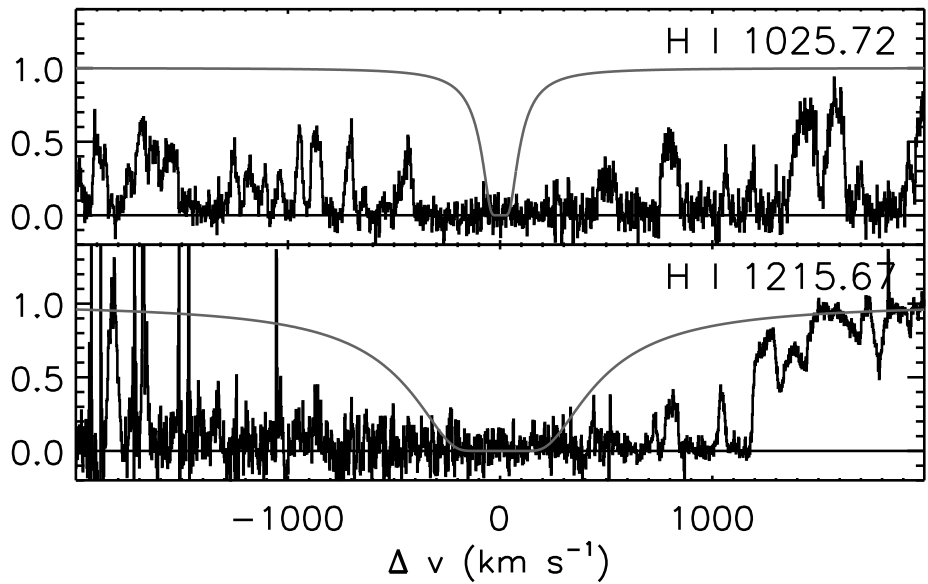}
  \figcaption{Voigt profile fit to \hi\ \lya\ and \lyb\ corresponding to
    the upper limit on $N_{\mhi}$ for the $z_{\rm sys} = 5.3364$ \oi\
    systems towards SDSS~J0231$-$0728.  In order to set an upper
    limit on $N_{\mhi}$ we have taken $b = 1$~\kms, although the
    actual $b$-parameter is likely to be significantly larger.  In
    this system, $N_{\mhi}$ is constrained by the transmitted flux in
    the wings of the \lya\ profile.}
  \label{hi_z5.3364}
\end{figure}

\clearpage

\begin{figure}
  \epsscale{0.75}
  \centering
  \plotone{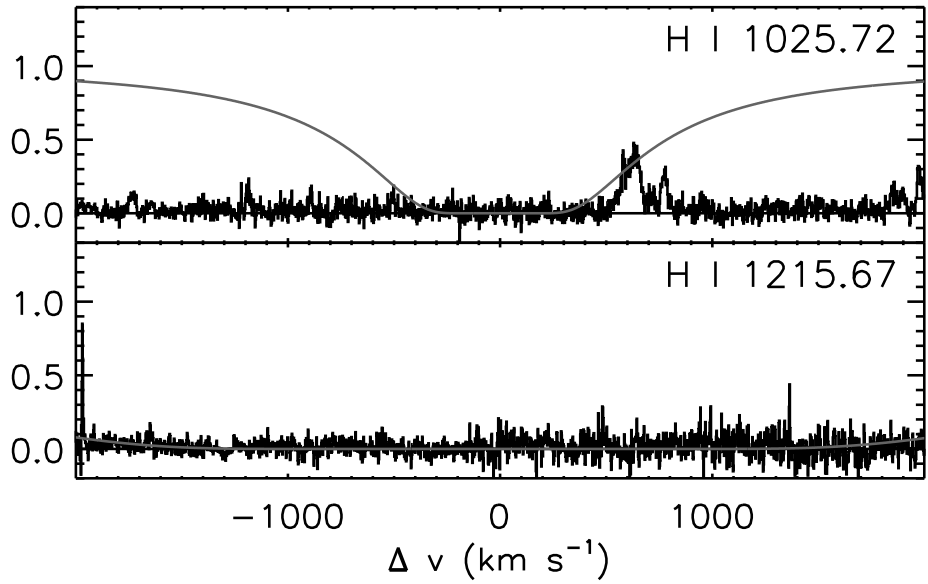}
  \figcaption{Voigt profile fit to \hi\ \lya\ and \lyb\ corresponding
    to the upper limit on $N_{\mhi}$ for the $z_{\rm sys} = 6.0097$
    \oi\ systems towards SDSS~J1148+5251.  In this system, $N_{\mhi}$
    is constrained by the transmitted flux in the red wing of the
    \lyb\ profile.}
  \label{hi_z6.0097}
\end{figure}

\clearpage

\begin{figure}
  \epsscale{0.9}
  \centering
  \plotone{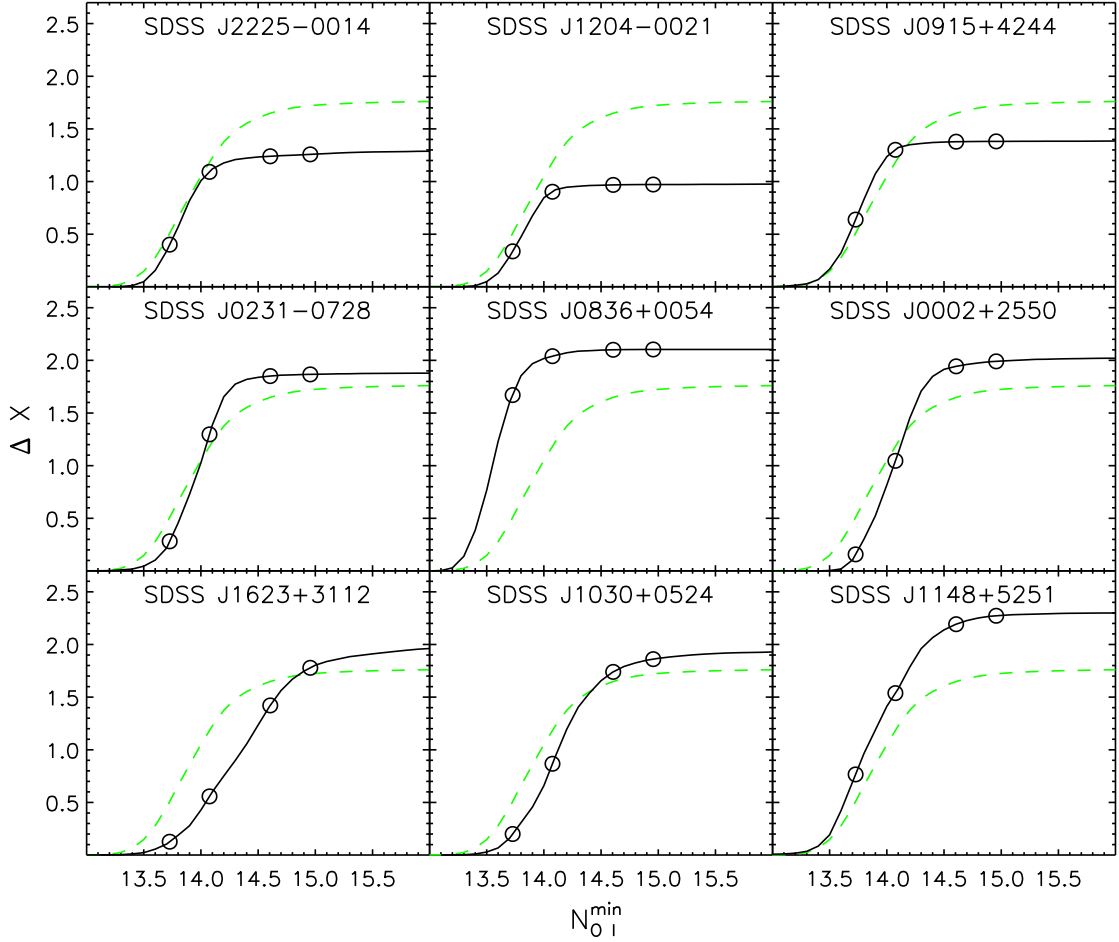}

  \figcaption{Sensitivity curves for each of our sightlines.  The
    \textit{solid lines} give the absorption pathlength interval over
    which we are sensitive to single-component \oi\ systems with
    $N_{\moi} \ge N_{\moi}^{\rm min}$ and relative abundances similar
    to those of the \oi\ systems in our sample.  \textit{Circles} mark
    the $\Delta X$ values corresponding to the $N_{\moi}$ of systems
    detected towards SDSS~1148+5251.  The \textit{dashed line} shows
    the mean $\Delta X$ as a function of $N_{\moi}^{\rm min}$ for all
    nine sightlines.  The three lowest-redshift sightlines have
    smaller maximum $\Delta X$ values due in part to $A$-band
    atmospheric absorption.}

  \label{sensitivity}
\end{figure}

\clearpage

\begin{figure}
  \epsscale{0.75}
  \centering
  \plotone{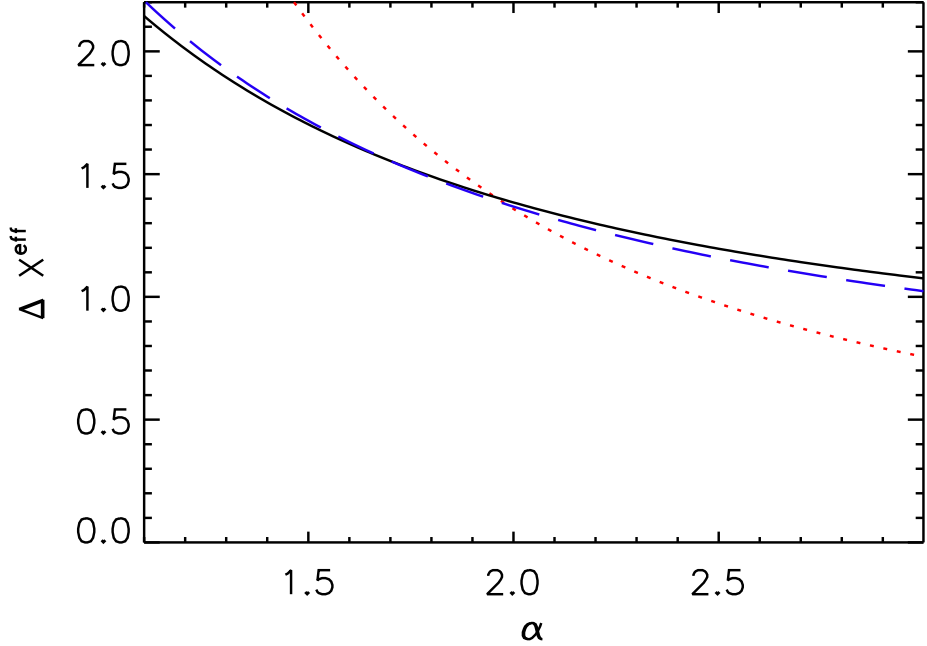}
  \figcaption{Effective absorption pathlength interval over which we
    are sensitive to \oi\ systems with $N_{\moi} \ge 10^{13.7}~{\rm
      cm^{-2}}$ as a function of $\alpha$, where $f(N_{\moi}) \propto
    N_{\moi}^{-\alpha}$ (see Eq.~\ref{eq:dxeffdef}).  The
    \textit{solid line} shows $\Delta X^{\rm eff}$ for SDSS~J1148+5251.
    The \textit{dashed line} shows the summed $\Delta X^{\rm eff}$ for the
    remaining eight sightlines, divided by 5.7 to match
    SDSS~J1148+5251 at $\alpha = 1.7$.  The \textit{dotted line}
    shows the combined $\Delta X^{\rm eff}$ for SDSS~1030+0524 and
    SDSS~J1623+3112.  For reasonable values of $\alpha$, the ratio of
    $\Delta X^{\rm eff}$ for SDSS~J1148+5251 and the remaining eight
    sightlines is essentially constant.  We would therefore expect to
    see roughly six times as many \oi\ systems along the other eight
    sightlines as towards SDSS~J1148+5251 for a non-evolving population
    of absorbers.}
  \label{dX}
\end{figure}

\clearpage

\begin{figure}
  \epsscale{0.75}
  \centering
  \plotone{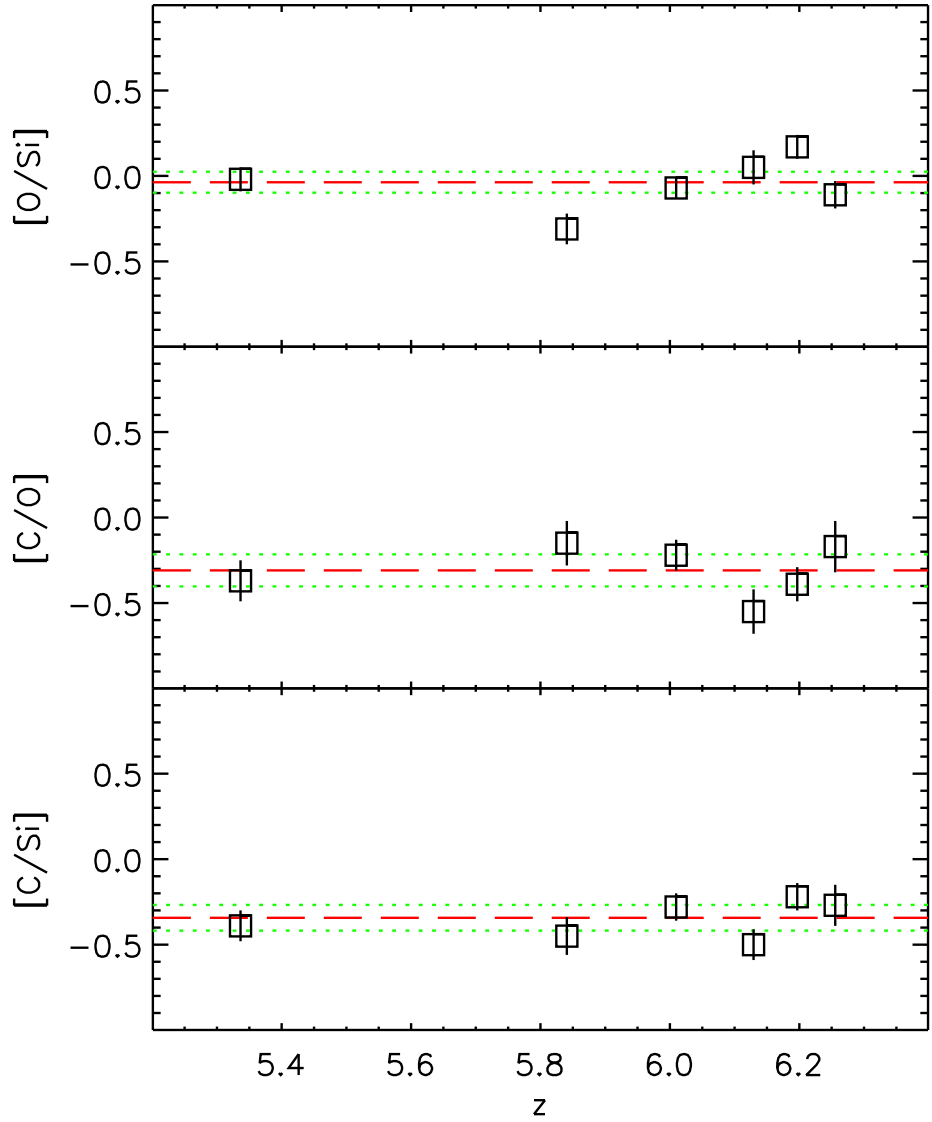}
  \figcaption{Relative metal abundances in the \oi\ systems assuming
    no ionization corrections.  Individual measurements plotted as
    \textit{squares} with $1\sigma$ error bars are taken from Table~3.
    \textit{Dashed lines} show the weighted mean values for all six
    systems.  \textit{Dotted lines} show the $2\sigma$ uncertainties
    in the means.}
  \label{relabundances}
\end{figure}

\clearpage

\end{document}